# Some Studies on k-essence Lagrangian

Somnath Mukherjee
Department of Physics, Dharsa Mihirlal Khan Institution(H.S),
G.I.P Colony, Howrah-711112, India.
Kalpana Chawla Centre for Space and Nano Sciences (KCCSNS), Kolkata700010, India
sompresi@gmail.com

**Abstract**

K-essence Lagrangian is studied in the context of an early universe epoch when t $\rightarrow 0$ . Equation of state parameter $\omega < -1$ as well as deceleration parameter $q_0 < -1$ , indicates an accelerated expansion of the universe at an early epoch of time driven by negative pressure generated by dark energy.

## 1. Introduction

An accelerated expansion of the universe has been confirmed by the observation of Type 1a Supernovae (SNe 1a) by The Supernova Cosmology Project [1-4] and the High-Z-Supernova search team [5-7]. Recent observation with Planck satellite [8] and WMAP satellite [9,10] led to the deep understanding of the mysterious energy known as dark energy as the source of negative pressure that leads to an accelerated expansion of the universe. These different observations conclude that 70% constituents of the universe are in the form of dark energy while 25% constituents are dark matter and the rest of about 5% constituents are observable in nature.

Several cosmological model has been established to understand the role of dark energy in the universe, out of which we have chosen k-essence model [11-15] of scalar field $\phi(r,t)$ with non-canonical kinetic term X as our field of study.

The k-essence scalar field $\phi(r,t)$ minimally coupled to background spacetime metric $g_{\mu\nu}$ has the action given by [16-22]

$$S_k[\phi, g_{\mu\nu}] = \int d^4x \sqrt{-g}\, K(\phi, X) \qquad (1)$$

where $K(\phi, X)$ is the k-essence Lagrangian and $X = \frac{1}{2} g^{\mu\nu} \nabla_\mu \phi \nabla_\nu \phi$ is the canonical kinetic term.

The energy -momentum tensor with respect to action (1) is given by

$$T_{\mu\nu} = \frac{-2}{\sqrt{-g}} \frac{\delta S_k}{\delta g^{\mu\nu}} = -K_X \nabla_\mu \phi \nabla_\nu \phi + g_{\mu\nu} K \qquad (2)$$

Where $K_X = \frac{dK}{dX}$ and $\nabla_\mu$ is the covariant derivative defined with respect to the gravitational metric $g_{\mu\nu}$ .

We will consider energy-momentum tensor of perfect fluid with density $\rho$ and pressure p

$$T_{\mu\nu} = (\rho + p) U_\mu U_\nu - p\, g_{\mu\nu} \qquad (3)$$

We will consider flat Friedmann-Lemaitre-Robertson-Walker (FLRW) metric of the form

$$dS^2 = dt^2 - a^2(t) \sum_{i=1}^{3} (dx^i)^2 \qquad (4)$$

where $a(t)$ is the cosmological scale factor.

We will choose the form of k-essence Lagrangian with non-canonical kinetic term [11-22] given by

$$K[\phi, X] = -V(\phi) F(X) \qquad (5)$$

where $V(\phi)$ is the scalar field potential and $F(X)$ is the kinetic part.
Comparing (2) and (3) energy density for the k-essence Lagrangian (5) is given by:-

$$\rho = V(\phi)[F(X) - 2XF_X] \qquad (6)$$

where $F_X = \frac{dF}{dX}$ .

## 2. Scaling relation

Equation of continuity is given by

$$\frac{d\rho}{dt}+3H(\rho+P)=0 \qquad (7)$$

where $H$ is the Hubble parameter defined in terms of scale factor $a(t)$ as

$$H=\frac{1}{a}\frac{da}{dt}=\frac{d\ln a}{dt} \quad .$$

From (6) we get

$$\frac{d\rho}{dt}=-V(\varphi)\left[F_X+2XF_{XX}\right]\frac{dX}{dt}+\frac{d\varphi}{dt}\left[F(X)-2XF_X\right]V_\varphi \qquad (8)$$

where $F_X=\frac{dF}{dX}$ , $F_{XX}=\frac{dF_x}{dX}$ and $V_\varphi=\frac{dV}{d\varphi}$ .

Putting (5), (6) and (8) in equation of continuity (7) we get

$$V(\varphi)\left[F_X+2XF_{XX}\right]\frac{dX}{dt}+\frac{d\varphi}{dt}\left[2XF_X-F(X)\right]V_\varphi+6HV(\varphi)XF_X=0 \qquad (9)$$

For a constant potential i.e., $V(\phi)$ =constant, so that $V_\phi=0,$ this reduces to an equation of form

$$\left[F_X+2XF_{XX}\right]\frac{dX}{dt}+6HXF_X=0 \qquad (10)$$

Since H= $\frac{1}{a}\frac{da}{dt}$ , this reduces to the form

$$\left[\frac{1}{X}+2\frac{F_{XX}}{F_X}\right]dX=-6\frac{da}{a} \qquad (11)$$

On integration this results in scaling relation of the form

$$\sqrt{X}F_X=Ca^{-3} \qquad (12)$$

where C is an integration constant. This is the scaling relation [12-14] which relates kinetic term with the cosmological scale factor a(t) of the FLRW metric. We will use this relation to determine k-essence Lagrangian for our study.

## 3. K-essence Lagrangian

Considering Friedmann equation

$$H^2=\frac{8\pi G}{3}\rho \qquad (13)$$

Comparing (6) and (13) we get

$$H^2=\frac{8\pi G}{3} \quad V\left[F(X)-2XF_X\right] \qquad (14)$$

Since we have considered constant potential hence we will write $V(\phi)=V=constant$ .

From (12) we get $F_X=\frac{Ca^{-3}}{\sqrt{X}}$ , substituting in (14) we get

$$F(X)=\frac{3}{8\pi GV}H^2+2C\sqrt{X}a^{-3} \qquad (15)$$

Let $q=\ln a$ , so that Hubble parameter becomes $H=\frac{\dot{a}}{a}=\dot{q}$ . Considering present observable universe to be homogeneous , we will consider scalar field to be function of time only i.e., $\phi(r,t)=\phi(t)$, so that $X=\frac{1}{2}\dot{\phi}^2$ . Equation (15) becomes

$$F(X)=\frac{3}{8\pi GV}\dot{q}^2+C\sqrt{2}\dot{\phi}e^{-3q} \qquad (16)$$

Putting (16) in (5) we get

$$K=-c_1\dot{q}^2-c_2\dot{\phi}e^{-3q} \qquad (17)$$

where $c_1=\frac{3}{8\pi G}$ and $c_2=C\sqrt{2}V$ .

This is the k-essence Lagrangian in canonical form with first term as the kinetic term and second term as a polynomial interaction. It has two generalised co-ordinate logarithm of the scale factor $q=\ln a$ and scalar field $\phi(t)$ . Interesting fact is logarithmic of the scale factor plays the role of a dynamical kinetic term in this Lagrangian. Setting up of Euler lagrangian equation and solution has been studied in Ref [14].

In this paper we will consider an early epoch and will try to develop k-essence theory of early universe so as to understand the nature of dark energy at the beginning of the universe.
At an early epoch Scale factor a(t) is very small so that q is also very small. Expanding 2nd term of (13) up to $q^2$ term we get

$$K=-c_1\dot{q}^2-c_2\dot{\phi}\left(1-3q+\frac{9}{2}q^2\right) \qquad (18)$$

Scaling $q \rightarrow q+\frac{1}{3}$ , equation (18) becomes

$$K=-c_1\dot{q}^2-\frac{9}{2}c_2\dot{\phi}q^2-\frac{1}{2}\dot{\phi} \qquad (19)$$

This Lagrangian with some modification has been studied in context to Ermakov invariant as an early probe to understand the universe in Ref [15]. Here in this work we will try to determine two most important cosmological parameters, Equation of state parameter $(\omega)$ and deceleration parameter $(q_0)$ to understand the role of dark energy at an early epoch of time.

**4. Equation of state parameter** $\omega$

Since lagrangian is equivalent to pressure , hence we can write the pressure (p) of k-essence field at a very early epoch when scale factor is very small as

$$p=-c_1\dot{q}^2-\frac{9}{2}c_2\dot{\phi}q^2-\frac{1}{2}\dot{\phi} \qquad (20)$$

The energy density is given by the Friedmann equation

$$\rho=\frac{3}{8\pi G}H^2=c_1\dot{q}^2 \qquad (21)$$

Equation of state parameter is defined as $\omega=\frac{p}{\rho}$ . From (20) and (21) we get

$$\omega=-1-\frac{\frac{9}{2}c_2\dot{\phi}q^2}{c_1\dot{q}^2}-\frac{\frac{1}{2}c_2\dot{\phi}}{c_1\dot{q}^2} \qquad (22)$$

Since $c_{1,}c_{2,}q^2$ and $\dot{q}^2$ are always positive , we see EOS $(\omega)$ remains negative if $\dot{\phi}$ remains positive i.e., if $\phi\sim t$ . This shows the scalar field is inflationary in nature. Thus equation (18) shows that $\omega<-1$ . This clearly indicates the presence of dark energy as a generator of negative pressure at an earlier epoch of time during the inflationary phase of evolution of the universe.

**5. Deceleration parameter $q_0$**

One of the important cosmological parameter is deceleration parameter $q_0 = \frac{-a\ddot{a}}{\dot{a}^2}$ , which determines the acceleration or deceleration of the expansion of the universe. It is defined in terms of EOS $(\omega)$ as

$$q_0 = \frac{1}{2}(1 + 3\omega) \tag{23}$$

Since from (18) we see that $\omega < -1$ at an early epoch then (23) shows that $q_0 < -1$ as well. Negativity of $q_0$ indicates an accelerated expansion of the universe at an earlier epoch of time.

**6. Conclusion**

In this paper we have studied the k-essence dark energy model under the context of an earlier epoch of time during the inflationary stage of evolution of the universe. Equation of state parameter $\omega < -1$ indicates the presence of dark energy as a source of negative pressure and deceleration parameter $q_0 < -1$ indicates an accelerated expansion of the universe during the earlier stages of time. Thus we get an accelerated expansion of the universe driven by negative pressure identified as dark energy at an earlier epoch of time.